\documentclass[sigconf, nonacm]{acmart}
\newcommand\vldbdoi{XX.XX/XXX.XX}

\newcommand\vldbavailabilityurl{URL_TO_YOUR_ARTIFACTS}
\usepackage{xspace}
\usepackage{multirow}
\usepackage{enumitem}
\usepackage{tcolorbox}
\usepackage{etoolbox}
\usepackage{alltt}
\AtBeginEnvironment{tcolorbox}{\footnotesize}
\tcbset{colback=blue!3,boxrule=0.03pt}
\usepackage{fvextra}
\usepackage{fancyvrb}
\fvset{commandchars=\\\{\}}
\usepackage{amsmath,amsfonts,bm}

\def\eqref#1{equation~\ref{#1}}
\def\1{\bm{1}}

\def\vx{{\bm{x}}}

\DeclareMathAlphabet{\mathsfit}{\encodingdefault}{\sfdefault}{m}{sl}
\SetMathAlphabet{\mathsfit}{bold}{\encodingdefault}{\sfdefault}{bx}{n}

\def\gL{{\mathcal{L}}}

\usepackage{hyperref}

\usepackage{flushend}

\newcommand{\para}[1]{\vspace{1.5mm}\noindent\textbf{#1.}}
\newcommand{\parait}[1]{\vspace{0mm}\noindent\textit{#1.}} %

\usepackage{todonotes}

\usepackage{bold-extra}

\newcommand{\modelname}{{{\textsc{CoddLLM}}}\xspace}

\newcommand{\claudetf}{Claude-3.5-Sonnet\xspace}

\newcommand{\texttotable}{\textsc{Text-to-Schema}\xspace}
\newcommand{\tabletotext}{\textsc{Row-to-Text}\xspace}
\newcommand{\mmlu}{\texttt{AnalyticsMMLU}\xspace}
\newcommand{\datadiscovery}{\textsc{Table Selection}\xspace}
\newcommand{\texttosql}{\textsc{Text-to-SQL}\xspace}

\newcommand{\fineweb}{\texttt{FineWeb-Edu}\xspace}
\newcommand{\birdselect}{\texttt{BIRD-TS}\xspace}
\newcommand{\openwikiselect}{\texttt{Open-WikiTable-TS}\xspace}
\newcommand{\wikipage}{\texttt{WikiPage-TS}\xspace}
\begin{document}

% \title{Enhancing Large Language Models for Data Analytics Applications}
\title{\modelname: Empowering Large Language Models for Data Analytics}


\author{Jiani Zhang$^1$ \quad Hengrui Zhang$^{2*}$
\quad Rishav Chakravarti$^1$  \quad  Yiqun Hu$^1$ \quad Patrick Ng$^1$ \\
Asterios Katsifodimos$^{1,3}$ \quad  Huzefa Rangwala$^1$ \quad George Karypis$^1$ \quad Alon Halevy$^{4*}$ }
\thanks{*Work done at AWS}
\affiliation{
\institution{$^1$Amazon Web Services, $^2$University of Illinois, Chicago, $^3$TU Delft, $^4$Google \\
\texttt{\{zhajiani,chakrris,yiqunhu,patricng,akatsifo,rhuzefa,gkarypis\}@amazon.com} \\ \texttt{hzhan55@uic.edu},~\texttt{halevy@google.com}
\country{}
}
}

\renewcommand{\shortauthors}{Zhang et al.}

\begin{abstract}
Large Language Models (LLMs) have the potential to revolutionize data analytics by simplifying tasks such as data discovery and SQL query synthesis through natural language interactions. This work serves as a pivotal first step toward the development of foundation models explicitly designed for data analytics applications. 
To propel this vision forward, we unveil a new data recipe for post-training LLMs, enhancing their comprehension of data management and empowering them to tackle complex real-world analytics tasks. Specifically, our innovative approach includes a scalable synthetic data generation method that enables the creation of a broad spectrum of topics centered on data representation and manipulation. Furthermore, we introduce two new tasks that seamlessly bridge tables and text. We show that such tasks can enhance models' understanding of schema creation and the nuanced translation between natural language and tabular data.
Leveraging this data recipe, we post-train a new foundation model, named \modelname, based on Mistral-NeMo-12B. To assess the language understanding and reasoning capabilities of LLMs in the realm of data analytics, we contribute \mmlu, a benchmark containing thousands of multiple-choice questions on databases, data analysis, and machine learning. Our focus on data discovery, has resulted in the contribution of three comprehensive benchmarks that address both database and data lake scenarios. 
\modelname not only excels in performance but also sets a new standard, achieving the highest average accuracy across eight datasets. It outperforms GPT-3.5-Turbo on \mmlu, exceeding GPT-4o by 12.1\% in table selection and showing an average improvement of 24.9\% in Text-to-SQL compared to the base model.

\end{abstract}

\maketitle

% \pagestyle{\vldbpagestyle}
% \begingroup\small\noindent\raggedright\textbf{PVLDB Reference Format:}\\
% \vldbauthors. \vldbtitle. PVLDB, \vldbvolume(\vldbissue): \vldbpages, \vldbyear.\\
% \href{https://doi.org/\vldbdoi}{doi:\vldbdoi}
% \endgroup
% \begingroup
% \renewcommand\thefootnote{}\footnote{\noindent
% This work is licensed under the Creative Commons BY-NC-ND 4.0 International License. Visit \url{https://creativecommons.org/licenses/by-nc-nd/4.0/} to view a copy of this license. For any use beyond those covered by this license, obtain permission by emailing \href{mailto:info@vldb.org}{info@vldb.org}. Copyright is held by the owner/author(s). Publication rights licensed to the VLDB Endowment. \\
% \raggedright Proceedings of the VLDB Endowment, Vol. \vldbvolume, No. \vldbissue\ %
% ISSN 2150-8097. \\
% \href{https://doi.org/\vldbdoi}{doi:\vldbdoi} \\
% }\addtocounter{footnote}{-1}\endgroup

% \ifdefempty{\vldbavailabilityurl}{}{
% \vspace{.3cm}
% \begingroup\small\noindent\raggedright\textbf{PVLDB Artifact Availability:}\\
% The source code, data, and/or other artifacts have been made available at \url{\vldbavailabilityurl}.
% \endgroup

\section{Introduction}
\label{sec:intro}

Large language models promise to usher in a new wave of innovation in data analytics~\citep{llm_data_wrangling2022,fernandez2023large,zhou2024llm,anderson2024design}. 
With LLMs, users will be spared the time-consuming tasks of discovering relevant data in messy data lakes, integrating diverse sources, and preparing the data for further use. Once the data is identified and prepared, users should be able to solve problems simply by asking questions in natural language, without having to navigate complex database schemas or domain-specific query languages. 
One method to realize these goals is by improving LLM performance on data-related tasks through prompt engineering~\citep{llm_data_wrangling2022,chorus}. 
However, this approach requires careful selection of optimal instructions and few-shot demonstrations for specific tasks, and it has not yet proven to be sufficient, often yielding erroneous answers~\citep{zhu2024chat2query}. In a related line of work,  models such as Table-GPT~\citep{table-gpt}, TableLlama~\citep{zhang2024tablellama}, TableLLM~\citep{wu2024tablebench}, and TableGPT2~\citep{su2024tablegpt2} have been finetuned with specific instructions for various table understanding and data wrangling tasks. Despite this progress, these works do not adequately address tasks that require a deep understanding of business concepts and their mapping to database schema and datasets. Existing works largely focus on tasks based on a single or a pair of tables and do not explore the relationships across various tables, which necessitates strong data modeling and integration capabilities.

To realize the vision mentioned above, we argue that LLMs must also be able to deal with messy collections of data, as often witnessed in data lakes. To do that, LLMs need to grasp a broad set of data management concepts. This includes understanding basic tabular representations, how business concepts are represented as complex database schemas or collections of interlinked datasets within a data lake, how tables can be created from other tables using views or other forms of computation, and the principles of data wrangling and integration~\citep{doan2012principles}. One key challenge is that models like GPT-4~\cite{achiam2023gpt} and Llama~\citep{touvron2023llama}, are primarily trained on general knowledge derived from web data, which limits their exposure to the specific training data from which they can learn fundamental data management concepts. Consequently, current LLMs perform poorly on tasks such as searching for relevant data within a large data lake that contains multiple interconnected datasets or evaluating hypotheses that require integrating data from diverse sources.

This work takes a first step toward developing foundation models that perform well on a broader set of data analytics tasks. 
We introduce a new data recipe designed for post-training any LLM, enhancing their ability to understand the "messy" reality of data management. More specifically, we post-train \modelname, a 12-billion-parameter foundation model based on Mistral-NeMo-Instruct, using our well-curated training corpus. Initially, we fine-tune \modelname on a large set of synthetically generated instruction-response pairs to enhance domain-specific knowledge. Next, we improve its data comprehension and problem-solving abilities by contributing two new table-text alignment tasks, followed by instruction fine-tuning on a smaller set of task-specific examples that focus on data discovery and real-world SQL code generation. Our \modelname demonstrates significant improvement over the base model and performs competitively with other state-of-the-art models, including GPT-4o, across various tasks, including three newly curated \mmlu datasets, three table discovery datasets, and two public \texttosql tasks.

\newpage
\noindent In this paper we make the following contributions:
\vspace{-3mm}
\begin{itemize}
    \item This work is the first to curate a large instructed training corpus designed for data analytics (summarized in Table~\ref{tbl:train-corpus}). The training corpus contains over 1 billion tokens and 2 million input-output examples.
    \item We propose a scalable synthetic data generation method that extracts and synthesizes instruction data from web corpora. This method grounds responses in reference documents to enhance diversity and avoid hallucination.
    \item We contribute two evaluation tasks and associated datasets. The first, named \mmlu, measures massive multitask language understanding in data analytics problems. The second, \wikipage is a human-annotated table selection benchmark with complex multi-table questions involving both textual and tabular data.
    \item We post-train a new 12B foundation model, \modelname, using the curated training corpus. 
    In the \mmlu evaluation, it surpasses GPT-3.5-Turbo, Mixtral-8x7B, and Mistral-Small. For \datadiscovery, it outperforms GPT-4o by 12.1\%. In the \texttosql task, it achieves an average improvement of 24.9\% compared to the base model.
    \vspace{-2mm}
\end{itemize}

\section{\modelname Overview}
In this section, we give an overview and motivation for how we curate data to train and evaluate \modelname for analytics tasks.

\para{Scalable Data Curation Methods}
Developing an expert-level LLM for data analytics is challenging due to the lack of high-quality, diverse, and supervised datasets. Human-annotated instruction datasets are limited in scale and can be expensive to obtain, while purely synthetic data often contains factual errors and lack diversity. To address these issues, we adopt an \emph{extraction-and-synthesis} strategy that leverages a large-scale web corpus rich in analytics-relevant knowledge and use cases. This approach involves identifying naturally occurring instruction data, such as question-answer pairs from Stack Overflow, and augmenting documents with instruction-response pairs grounded in their content. After content distillation, we remove the plain text documents to form the final dataset. It is worth noting that by grounding responses in reference documents during the generation process, we can reduce hallucinations and increase the diversity of the dataset.  Empirical studies show that post-training with instruction-aware data enhances model performance more than plain text, as it aligns the model better with domain-specific queries and responses~\citep{chung2024scaling,instruction_pre_training}.

\para{New Training Tasks}
To improve the model's understanding of the relationship between natural language and tabular data, we design two table and text alignment tasks. The first task, \texttotable, is to generate a table schema from textual scenario description, which allows the to model understand how different pieces of information relate to, and are expressed with, various business entities. The second task, \tabletotext, aims to generate a text description for every row in a table. Mastering this task is essential for enhancing the model's ability to translate structured information into human-readable formats and vice versa, which is crucial for tasks such as generating reports, summaries, and data motivations.

\begin{table}[t!] 
 \centering
 \caption{Training Corpus Overview.
 \vspace{-4mm}
 } 
 \label{tbl:train-corpus}
 \resizebox{\columnwidth}{!}
 {
    \begin{tabular}{llrrc}
        \toprule[1.2pt]
        & \textbf{Tasks (per chapter)} & \begin{tabular}[c]{@{}r@{}} \#Documents/ \\ \#Examples \end{tabular} &  \#Tokens & Source  \\
        \midrule
        \multicolumn{4}{l}{\hspace{-.5em} \textit{Chapter1: Knowledge Corpus}} \\
        \cmidrule{2-5}
        & Plain Text & 5.8M & 1.9B & Web\\ \cmidrule{3-5}
        & Instruction-Response Pairs & 8.8M & 0.9B & \begin{tabular}[c]{@{}r@{}}Web \& \\ Synthetic \end{tabular}  \\
        \midrule
        \multicolumn{4}{l}{\hspace{-.5em} \textit{Chapter2: Text\&Table Alignment}} \\
        \cmidrule{2-5}
        & \texttotable & 4.8K  & 7.4M  & Synthetic\\
        & \tabletotext & 53.6K & 12.3M & Synthetic \\
        \midrule
        \multicolumn{4}{l}{\hspace{-.5em} \textit{Chapter3: Analytics Tasks}} \\
        \cmidrule{2-5}
        & \datadiscovery & 17.4K   & 37.8M  & Public \\
        & \texttosql     & 121.8K  & 42.6M  & Public \\
        \bottomrule[1.2pt]
        \multicolumn{2}{l}{\hspace{.5em} Total} & 9.0M & 1.0B
    \end{tabular}
}
\vspace{-3mm}
\end{table}

\para{A Large-scale Training Corpus}
The curated data consists of approximately 2 million examples and 1 billion training tokens. We organize this training corpus into distinct "chapters", much like a well-structured textbook.
\textit{Chapter 1} covers knowledge in data management and analysis, including data modeling concepts, instructions for using analysis tools, machine learning techniques, and more. We prepare this chapter by filtering relevant content from the \fineweb dataset~\citep{penedo2024fineweb} and applying the extraction-and-synthesis method to automatically generate instruction pairs.
\textit{Chapter 2} focuses on the foundational schema-level and row-level table and text alignment tasks, as mentioned above.
\textit{Chapter 3} addresses specific analytics tasks, particularly those that require identifying and integrating data from multiple sources. We chose \datadiscovery and \texttosql.
The goal of \datadiscovery is to identify one or more datasets that contain the necessary information to answer a user-specified question. \texttosql is a well-established task that involves translating natural language instructions into executable SQL queries. 

\para{New Evaluation Tasks and Datasets}
To evaluate the foundation model’s performance, we introduce \mmlu, a new benchmark with three datasets designed to assess the model's language understanding and reasoning capabilities in the analytics domain. This benchmark consists of thousands of multiple-choice questions covering the areas of database management, data analysis, and machine learning. We collect questions from existing textbooks and synthetically generate additional question-answer pairs with \claudetf. All answers are manually reviewed and revised. 
Additionally, we prepare three new datasets for the \datadiscovery (TS) task. The key challenge in this task is that the model needs to determine how many tables are required in the answer. Three datasets cover distinct settings:
\birdselect -- derived from BIRD~\citep{bird}, where candidate tables come from a single database with well-structured schemas, 
\openwikiselect -- sourced from Open-WikiTable~\citep{open_wikitable}, where the candidate tables are semantically similar, 
and \wikipage -- a newly annotated benchmark, which draws input from a single Wikipedia page containing multiple tables and text descriptions on the same topic. For \wikipage, we aim at questions that require multi-hop reasoning, in which models must integrate information from multiple tables or perform sequential logical steps to derive the correct answer.

\subsection{Summary of Evaluation Results}
We evaluate \modelname on three key aspects: knowledge testing, data selection, and SQL code generation. Overall, \modelname shows significant improvement over the base Mistral-NeMo model and delivers competitive performance compared to the leading models.
\begin{itemize}
    \item \modelname achieves the highest overall score of 0.697, making it the best-performing model.
    \item In the \mmlu evaluation: \modelname surpasses most other models, including GPT-3.5 Turbo and Mixtral-8x7B, although it falls relatively 5.8\% short of GPT-4o.
    \item For \datadiscovery, \modelname outperforms GPT-4o by 12.1\%, making it the most potent model for this task. 
    \item   In the \texttosql evaluation, \modelname achieves an average execution accuracy of 0.576, outperforming all other models and demonstrating an average of 24.9\% relative improvement compared to the base model.
\end{itemize}

\section{Preliminaries}
\noindent \textbf{Autoregressive Language Modeling}.
Language provides a versatile way to represent tasks, data, inputs, and outputs, all as a sequence of tokens. Autoregressive language modeling is the basis for LLMs like GPT~\citep{gpt2,gpt3}. This approach predicts the probability of a sequence of words or tokens, with each prediction conditioned on the previous elements in the sequence. 

Formally, given a language token sequence $\vx = (x_1, x_2, \cdots, x_n)$, autoregressive language modeling decomposes the joint distribution of the sequence as the product of a series of conditional probabilities: $p(\vx) = \prod_{i=1}^n p(x_i|x_1, ..., x_{i-1})$,
where $p(x_1|x_0) = p(x_1)$ is the marginal probability. With the factorized distribution and a parameterized model (e.g., Transformers~\cite{vaswani2017attention}), the parameterized distribution $p_{\theta}(x)$ can be optimized via minimizing the negative log-likelihood loss:
\begin{equation}\label{eqn:loss-ar}
    \gL(\theta) = -\log p_{\theta}(\vx) = - \sum\limits_{i=1}^n \log p(x_i|x_{1},\cdots, x_{i-1}).
\end{equation}

\vspace{3pt}
\noindent \textbf{Query-based Data Analytic Tasks}. This paper focuses on query-based data analytic tasks, represented as $\{task, data, query, answer\}$. Given a task description, the data to be analyzed, and a natural language query, the goal is to predict the answer, i,e, $p(answer|task,$ $data, query)$. For example, we can format the input text for a table selection task as: "$\texttt{find tables},$ $\texttt{table schema},~$ $\texttt{who was the}$ $\texttt{only athlete...}$". Then, the model is expected to return the table name(s) that can answer the question, e.g., "$\texttt{<tables>Final\_1},$ $\texttt{Athletics\_1</tables>}$".

\vspace{3pt}

\noindent \textbf{Supervised Instruction Tuning} is a critical fine-tuning process employed to enhance the performance of LLMs on specific tasks by leveraging labeled datasets. Supervised instruction tuning focuses on adapting the model to follow explicit instructions and produce task-specific outputs. This process involves training the model on input-output pairs, where the inputs are typically natural language instructions or prompts, and the outputs are the desired responses or completions. The loss function of supervised instruction tuning is computed only on the "output" tokens to optimize the ability to execute specific tasks and understand instructions.
\section{Training Corpus Creation}\label{sec:data_curation}
We curate a large and diverse collection of analytics-specific training corpus with three specific objectives: (1) ensuring broad coverage of various data analytics knowledge concepts, (2) improving the model's comprehension of database schema and table content, and (3) incorporating task-specific input-output examples to enhance the model's capabilities in solving real-world analytics problems.

The training corpus is formulated as akin to a textbook, presenting each data component as a distinct chapter. 
\emph{Chapter 1} contains vast knowledge filtered from web-crawled data using a purposely trained classifier. These knowledge elements are formatted into the input-output pairs to enhance the model’s learning efficiency.
\emph{Chapter 2} expands on this foundational stack by introducing two new tasks: 1) \texttotable, which involves designing a schema based on a provided scenario description, and 2) \tabletotext, which generates a text description for each row in a given table. These tasks enhance the model's ability to understand and translate between the modalities of natural language text and tabular data.
\emph{Chapter 3} includes \datadiscovery and
\texttosql as representative downstream analytics tasks of the real world.

\subsection{Chapter 1 - Analytics-specific Knowledge} 
\label{sec:chapter1}

\begin{figure}[tb]
 \centering
 \includegraphics[width=\linewidth]{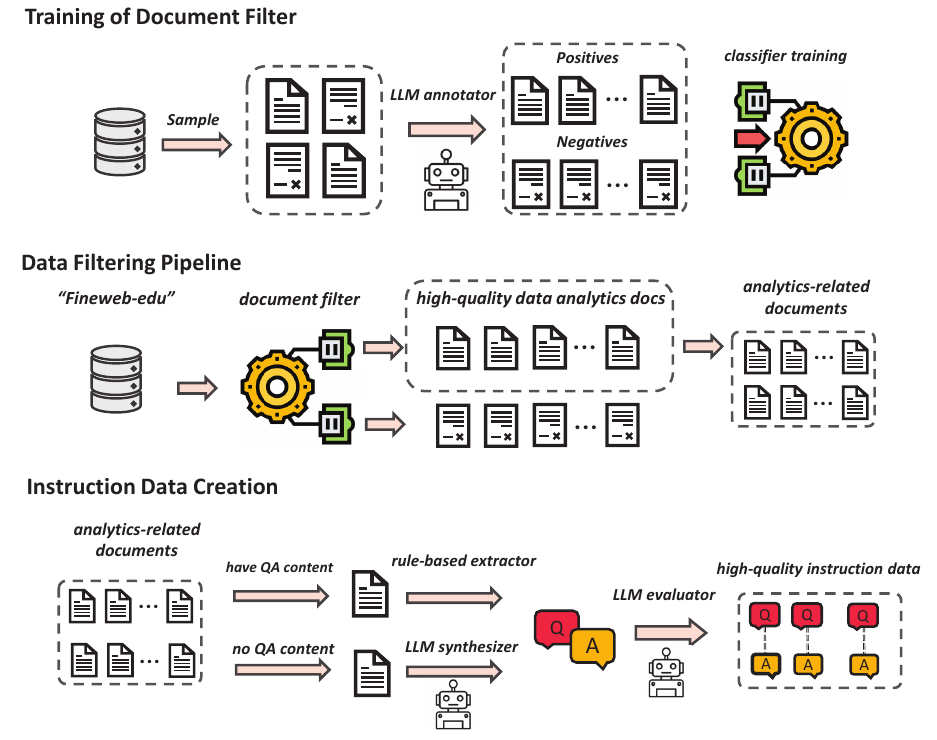}
 \caption{Building Chapter 1 data. Step 1 (the top
and middle figures): We first train a document classifier using annotations from a LLM. After training, we use the classifier to filter documents related to data analytics from the \fineweb dataset. Step 2 and Step 3 (the bottom figure): The filtered documents are then converted into question-answer pairs using a rule-based extractor or a LLM-based synthesizer. Finally, we adopt LLM-as-a-judge to eliminate low-quality examples.}
 \label{fig:chapt1-pipeline}
 \vspace{-2mm}
\end{figure}

Training a domain-specific LLM requires a vast corpus of relevant knowledge. However, well-organized datasets on data analytics are still scarce. Recent studies have explored the use of seed data to prompt LLMs in order to expand domain-specific datasets~\citep{amini2019mathqa,zhou2024llm,yumetamath}. While this approach shows promise, synthesized data that lacks proper references often exhibits significant biases, lacks diversity, and is prone to hallucinations~\cite {yue2024mammoth2}.
To address these challenges, we use a three-step pipeline to curate large-scale analytics-related instruction-response pairs (see Figure ~\ref{fig:chapt1-pipeline}): 
\begin{enumerate}
    \item \textbf{Filtering:} Identifying and extracting data analytics-related documents -- such as finance analysis, sales prediction, and code-related data -- from large-scale web sources;
    \item \textbf{Instruction Creation}: Converting  plain text documents into question-answer pairs via extraction and synthesis;
    \item \textbf{Assessment:} Evaluating the extracted QA pairs to eliminate low-scored examples and enhance dataset quality.
\end{enumerate}

\subsubsection{Step 1: Filtering Analytics-specific Documents}
We use the \fineweb dataset~\citep{lozhkov2024fineweb-edu} as our source, which contains 1.3 trillion tokens of educational-quality web pages. Due to the dataset's multi-domain content, manually identifying data analytics-related documents presents significant challenges. To address this, we develop a model-based filtering approach to extract relevant documents. This process consists of three stages: a) \emph{LLM as a document rater}, by leveraging LLMs to assess document relevance in order to obtain labels; b) \emph{Training a content classifier}, by developing a supervised model based on rated samples; c) \emph{Deploying the trained filter} by applying the classifier to extract analytics-related documents. We now detail these stages.

\para{LLM as a Document Rater} 
Training the document classifier requires both positive samples (high-quality documents relevant to data analytics) and negative samples (low-quality or minimally relevant documents). 
Inspired by the effectiveness of LLM-as-a-judge in automatic text evaluation~\citep{gunasekar2023textbooks}, we sample approximately 0.1\% of the \fineweb dataset and use \claudetf to score each document’s relevance to data analytics on a scale of 1 to 5. 

The following example illustrates how \claudetf evaluates a document based on its textual quality and relevance to data analytics.
The document consists of a brief explanation of SQL statements, describing their structure and syntax. The rater assigns a score of 5, indicating high relevance to SQL and data analytics.

\begin{tcolorbox}[left=1pt, right=0pt, top=1pt, bottom=1pt]
\begin{verbatim}
# Document: 
SQL statements have a specific structure and syntax that must be 
followed precisely. A basic SQL statement is composed of several 
Clauses:
```sql
SELECT column1, column2
FROM table_name
WHERE condition;
```

# Rater Output:
This text clearly focuses on SQL, which is a fundamental tool in 
data analytics...
Score: 5 
\end{verbatim}
\end{tcolorbox}

\para{Training a Document Classifier} After obtaining the annotated data, we train a Transformer-based text classification model. Considering that most documents are quite long -- some exceed 2,000 tokens, we build the classifier on top of LongFormer~\citep{beltagy2020longformer}, a Transformer model supporting a maximum context length of 4,096 tokens. The transformer categorizes the documents into five classes according to the annotated scores (1 to 5). After training, the classifier achieves 78\% test accuracy and 87\% recall in retrieving high-rating documents, demonstrating its effectiveness in filtering analytics-related content from the raw corpus.

\para{Deploying the trained filter on \fineweb} We then apply the trained classifier to filter documents in the \fineweb dataset, selecting those with a score of $4$ or higher.
This filtering process reduces the dataset to 4.9\% of its original size, consisting of over 5.8 million documents and 1.9 billion tokens.
The following passages present an example document from \fineweb, which provides a clear and accurate introduction to the concept of the "foreign key", a crucial concept in databases. Due to its relevance to data analytics, it is classified as "score = 5" and included in Chapter 1.

\begin{tcolorbox}[left=1pt, right=0pt, top=1pt, bottom=1pt]
\begin{Verbatim}
# An example of a filtered document (score = 5)

In database design, the concept of relationships between tables 
is crucial for maintaining data integrity and ensuring efficient
data retrieval. When two tables need to be linked, a special type 
of attribute is used to establish this connection. This attribute, 
known as a foreign key, serves as a reference to the primary key
of another table.
\end{Verbatim}
\end{tcolorbox} 

\subsubsection{Step 2: Instruction Extraction/Synthesis}
Recent studies have demonstrated the effectiveness of instruction data in post-training LLMs, showing that LLMs perform better when trained on instruction-response formatted data rather than plain text with the same semantic meaning~\citep{instruction_pre_training}. Motivated by the extraction and refinement strategy for extracting mathematical contents from raw documents~\citep{yue2024mammoth2}, we propose an \emph{extraction and synthesis} strategy. For each filtered document, we either extract native question-answer pairs using predefined rules or generate synthetic question-answer pairs by grounding them in the documents and ensuring explicit references to the content. The goal is to construct a synthetic instructed version through content distillation and then remove the plain text documents to form the final dataset.

\para{Extraction} We first use regular expressions to identify potential indicators of QA pairs -- "[Qq]uestion:" and "[Aa]nswer:" -- to classify the 5.8M collected analytics-related documents. Among these, about 2.8K documents contained both question and answer keywords. We then leverage \claudetf to extract question-answer pairs from these documents, yielding a total of 46K QA pairs, as a single document might contain multiple QA pairs. After conducting a pilot study with two human annotators on 300 samples, we observed that 97\% of the extracted pairs were deemed valid and, therefore, can be used as our training corpus.

\para{Synthesis} For the remaining documents where explicit question-answer pairs cannot be extracted, we prompt \claudetf to synthesize question-answer pairs of varying difficulty levels, ensuring the generation of relevant QA pairs of high quality. Specifically, we require that the generated QA pairs meet the following criteria: 1) Varying difficulty: Include questions ranging from basic common sense to advanced data-related topics. 2) Context-dependency: Ensure simple questions rely on common knowledge, while complex ones require provided context for answers. 3) Diverse format: Use varied question types beyond "How" and "What", and encourage longer, detailed questions and answers where possible. Following this QA generation pipeline, we obtain about 8.8M QA pairs. Below is an example of the synthesized question-answer pair and the original document content, where the original document is a passage about the two outputs from a forecasting operation in Excel.

\begin{tcolorbox}[left=1pt, right=0pt, top=1pt, bottom=1pt]
\begin{Verbatim}

# Original document:
"There are two outputs generated from the forecasting operation. 
Firstly, \textbf{a worksheet is added to the workbook, which shows the} 
\textbf{projected values as a dashed line}. Secondly, \textbf{the projected values} 
\textbf{are added sequentially to the original table.} It should be noted 
that the chart is actually generated from the original table and 
the forecasts that were appended as part of data mining operation."

# A question-answer pair
"question": "What are the two outputs generated from  the 
forecasting operation in the Excel add-in?"
"answer": "The two outputs generated from the forecasting 
operation in the Excel add-in are: 1)\textbf{ a worksheet added to the} 
\textbf{workbook showing the projected values as a dashed line}, and 2) 
\textbf{the projected values appended sequentially to the original table.}"
\end{Verbatim}
\end{tcolorbox}

\subsubsection{Step 3: Assessment}
After obtaining all QA pairs, we employ a widely-used LLM-as-a-Judge, \texttt{Prometheus-eval}~\citep{kim2024prometheus}, to conduct two rounds of filtering for each QA pair, further improving the quality of Chapter 1 data. In the first round of screening, we focus on the completeness of Questions and Answers, ensuring that all information necessary to arrive at the answer is contained within the Question text. In the second round of screening, we further evaluate the accuracy of the answers, ensuring that only correct question-answer pairs are retained in the training dataset. Eventually, we curate a dataset of 8.8 million instruction-response pairs, comprising a total of 0.9 billion training tokens.

\subsection{Chapter 2 - Table and Text Alignment}
\label{sec:chapter2}
In this chapter of the textbook, we design two tasks aimed at enhancing the LLM's ability to understand and process both tabular and textual data. These tasks are designed to facilitate the model's ability to translate between textual and structured data modalities.

\subsubsection{\texttotable}
The objective of this task is to generate a table schema from an input description of a database or information system (e.g., a credit card transaction system). Training on this task enables the model to understand how different pieces of information relate to each other within a structured representation. The underlying hypothesis here is that by learning to design schemas, the model will also develop the ability to interpret schemas when it receives them as inputs~\citep{schemamatch}.

The main challenge in acquiring scenario descriptions and schema pairs stems from their infrequent co-occurrence within a single database or document. To tackle this issue, we generate synthetic pairs by translating well-structured schemas into multiple scenario descriptions using LLMs. The data generation steps begin with filtering 4,399 high-quality schemas from the overall 221,171 schemas in SchemaPile~\citep{schemapile}. We use \claudetf to assess the quality of the data. After classification, three different scenario descriptions are generated for each schema. These generated descriptions are then evaluated using Prometheus-eval~\citep{kim2024prometheus} based on the criterion, ``Is the scenario description concise and relevant?''. Only descriptions that receive a score higher than four are retained. As a result, we generate 4.8 thousand examples (refer to Table~\ref{tbl:train-corpus}).

\subsubsection{Row-to-Text} 

Given a table and contextual data, the goal is to generate a detailed text description for a given row in the table. Training a model on a large number of \tabletotext examples improves its ability to convert structured tabular data into meaningful, fluent, and accurate text. Below, we present an input-output example from the sports domain. The input includes a task instruction and a single row from a table, along with the page title, section title, and caption. The output is a detailed text description. 

\begin{tcolorbox}[left=1pt, right=0pt, top=1pt, bottom=1pt]
\begin{Verbatim}
# Input:
Write a detailed description for the row in the table.
Enclose the description in <row_description> tags.

Page Title: Badminton at the Pan American Games
Section Title: Medal table
Caption: Medal table
| Rank |    Nation     | Gold | Silver |  Bronze | Total |
| ---- | ------------- | ---- | ------ | ------- | ----- |
\textbf{|  1   | Canada        |  16  |   16   |    11   |   43  |}

# Output:
<row_description>
\textbf{Canada} won \textbf{16} gold medals, \textbf{16} silver medals, and \textbf{11} bronze medals, 
totaling \textbf{43} medals in badminton events at the Pan American Games,
ranking \textbf{first} among all participating nations.
</row_description>
\end{Verbatim}
\end{tcolorbox}

The key idea behind the data generation process is to take real-world tables as input, leverage LLMs to generate row descriptions, and then apply a filtering process to keep high-scored examples. We utilize tens of thousands of WikiTables from the open-wikitable dataset~\citep{open_wikitable}. \claudetf is employed to generate descriptions for each row within these tables. To ensure the quality and accuracy of these descriptions, we use an entailment classifier to filter tables, keeping only those where all row descriptions were classified as either entailment or neutral. This filtering process guarantees that each generated description aligns with the original table entries. Overall, we generate 46.3 thousand \tabletotext pairs and 10.9 million tokens (refer to Table~\ref{tbl:train-corpus} for additional data statistics).

\subsection{Chapter 3 - Data Analytics Tasks}
\label{sec:chapter3}
Chapter 3 presents carefully curated training examples designed to enhance model performance on downstream analytics tasks, with a focus on real-world applications. We prioritize two critical tasks: \datadiscovery and \texttosql conversion. These foundational tasks are selected to provide our model with a comprehensive understanding of datasets, enabling and augmenting their capacity to execute more complex analytics tasks.

\subsubsection{\datadiscovery}
\label{sec:chapter3_table_selection}
Relevant data selection from large collections of datasets is a long-standing challenge that organizations struggle with to this day~\citep{hulsebos2024took}. In this task, we focus on \textit{question-based search}, where the goal is to identify one or more data pieces that contain the necessary information to answer a user-specified question. The data can be structured or unstructured, often requiring integration from multiple sources. In this context, the ability to recognize relationships between datasets can significantly enhance both the accuracy and efficiency of task completion.

To create training examples, we use the training set of BIRD and Open-WikiTable and convert these datasets to serve the needs of the \datadiscovery task.  Section~\ref{sec:table_selection_prompts} gives details on task examples.

\para{Leveraging BIRD for \datadiscovery} 
The BIRD dataset~\citep{bird} was originally designed for text-to-SQL tasks, where each example includes a question, multiple tables from a database, and a corresponding ground truth SQL query. We select 8,954 questions from its training set and include all tables from the same database as potential candidates. We use only the schemas to highlight table relationships, without using any table contents. We identify the ground truth tables from their associated gold SQL queries.

\para{Leveraging Open-WikiTable for \datadiscovery} The Open-The WikiTable~\cite{open_wikitable} dataset was designed for open-domain question answering over tables. Each question is paired with its corresponding reference ground truth table. For each question, we use BM25~\citep{bm25}, a term-based sparse encoding ranking algorithm, to retrieve ten tables from a pool of 24,680 WikiTables as the candidate table set. These ten retrieved tables are ranked based on their relevance to the query, with each table sharing some degree of term-based similarity to the question. However, the retrieved tables may not be directly related to one another. 
We retain 8,484 questions from the training set and ensure that the retrieved ten tables include the gold tables. We provide metadata for the WikiTables, including page titles, section titles, and captions, if available. To ensure scalability for handling large tables, we present only three sampled rows from each table to the LLMs. 

\vspace{-1mm}
\subsubsection{\texttosql}

\texttosql is a well-established data analytical task that aims to convert natural language questions into SQL queries, extracting desired information with proper computation and transformation performed on relational databases. The challenge lies in accurately interpreting the semantics of the input text, mapping it to the appropriate database schema, and generating a syntactically correct SQL query that fulfills the user's intent. 
Unlike traditional methods that rely on predefined templates or rules\citep{li2014constructing}, LLMs have become the predominant tools for this task, as highlighted in recent studies~\citep{gao2024xiyan, pourreza2024chase, talaei2024chess}. Common approaches with LLMs include prompt engineering and task-specific fine-tuning. 
Prompt engineering employs techniques such as few-shot learning and multi-step reasoning without updating the model weight, but sometimes producing incorrect results for complex SQL queries. On the other hand, fine-tuning enhances performance by training the LLM on task-specific data, but it compromises the model's general instruction-following ability. In contrast, \modelname takes a different approach by offering a domain expert model that excels not only in \texttosql but also across other analytical tasks.

We extract a sample of approximately 9K examples from the BIRD~\citep{bird} dataset's training set and 7K examples from the Spider~\citep{yu-etal-2018-spider} dataset's training set. Additionally, we incorporate 105.8K examples from synthetic text-to-SQL~\citep{gretel-synthetic-text-to-sql-2024}.
For each example, we construct the input text prompt by concatenating three components: \textsc{<Schema> <Task Instruction> <Question>}. To represent the schema, we use \textsc{CREATE TABLE} SQL statements for the tables. The output is the SQL query enclosed within the \texttt{<SQL>} tags. For more data processing details, see Section~\ref{sec:exp_text2sql}.

\subsection{Model Architecture}

\modelname is built on top of Mistral-NeMo-Instruct, a decoder-only Transformer of 12B parameters, with the following specifications:
\begin{itemize}
    \item 128k vocabulary size
    \item Rotary Positional Embedding (RoPE)~\citep{su2024roformer}
    \item 40 Transformer layers~\citep{vaswani2017attention}
    \item 5,120 token dimension and 14,436 hidden dimensions
    \item 128 head dimensions
    \item 32 attention heads and 8 grouped query attention~\citep{ainslie2023gqa}
    \item SwiGAU activation~\citep{shazeer2020glu}
\end{itemize}

\subsection{\modelname Training}

We train \modelname using the Nvidia NeMo framework, leveraging its capabilities for efficient distributed training with data and model parallelism techniques. 

We begin the \textit{supervised instruction-tuning} process using input-output pairs from Chapter 1 (see Section~\ref{sec:chapter1}). By focusing on the basic knowledge of data analytics, the model can establish a strong understanding of key concepts, terminology, and reasoning patterns specific to data analytics. This foundational knowledge is crucial before the model addresses more complex, task-specific scenarios. We train the Chapter 1 data for one epoch. As noted by~\citet{hernandez2022scaling}, repeated tokens can negatively impact performance.

Next, we continue with supervised fine-tuning using data from Chapters 2 and 3 to improve the model's generalization capabilities, enabling it to apply its foundational knowledge to diverse downstream tasks more effectively. To ensure consistency across datasets, we standardize the format of these chapters in our analytic data corpus, aligning them with a unified data structure. This standardization ensures that the fine-tuned models can process data uniformly, regardless of the original chapter formats, thereby streamlining the training process and enhancing its efficiency.

\para{Training Configurations}
We use the AdamW~\citep{loshchilov2017fixing} optimizer with $\beta_1 = 0.9, \beta_2 = 0.98, \epsilon = 10^{-8}$, and a weight decay of $0.01$. We set the maximum learning rate to $lr_{\text{max}} = 1e^{-6}$, the minimum learning rate to $lr_{\text{min}} = 0$, with a Cosine learning rate scheduler to allow the model to make fine-grained adjustments with the labeled data. 

\para{How to choose the \modelname checkpoint?} We train the Chapter 1 data for one epoch, followed by training on Chapter 2 and Chapter 3 data for two epochs while mixing 10\% uniformly selected Chapter 1 data. The final endpoint is selected as \modelname for evaluation in all the following experiments. 
\section{Evaluation Tasks and Experiments}

We evaluate \modelname's reasoning capabilities across various analytics tasks, mainly focusing on domain-knowledge testing, table selection, and text-to-SQL tasks.

\para{LLM Baselines} We compare \modelname with five open-source LLMs, all obtained from Hugging Face: \href{https://huggingface.co/mistralai/Mistral-7B-Instruct-v0.3}{\texttt{Mistral-7B-Instruct-v0.3}}, \href{https://huggingface.co/mistralai/Codestral-22B-v0.1}{\texttt{Codestral-22B-v0.1}}, \href{https://huggingface.co/mistralai/Mistral-Small-Instruct-2409}{\texttt{Mistral-Small-Instruct-2409}}, \href{https://huggingface.co/mistralai/Mixtral-8x7B-Instruct-v0.1}{\texttt{Mixtral-} \\ \texttt{8x7B-Instruct-v0.1}}, as well as the base model \href{https://huggingface.co/mistralai/Mistral-Nemo-Instruct-2407}{\texttt{Mistral-Nemo-} \\ \texttt{Instruct-2407}}. We use the instruction-tuned versions for all baseline models that have been fine-tuned on extensive general-purpose tasks. Meanwhile, we include three closed-source OpenAI models: \texttt{GPT-3.5-Turbo}, \texttt{GPT-4o-mini}, and \texttt{GPT-4o}, establishing strong baselines. Additionally, we included some task-specific baselines where appropriate. Unless explicitly specified, we report results on \modelname and all the baselines in a zero-shot setting, with no demonstration examples provided during inference.

\begin{table}[t!] 
 \centering
 \caption{Overview of all the evaluation benchmarks. "Exec. Acc." refers to Execution Accuracy. "MCQ" stands for Multiple Choice Questions. We newly introduce three \mmlu datasets and a human-annotated \wikipage dataset. Our training corpus does not include any subset of these four datasets, so we consider them "unseen" datasets.} 
 \label{tbl:all_eval_dataset}
 \vspace{-3mm}
 \resizebox{\columnwidth}{!}
 {
    \begin{tabular}{llccccc c c}
        \toprule[1.2pt]
        & \textbf{Dataset} &  \#Example  & Metric & Sources &  \begin{tabular}[l]{@{}l@{}} Seen/Unseen \\ In Training \end{tabular}  \\
        \midrule
        \multicolumn{3}{l}{\hspace{-.5em} \textit{\mmlu}} \\
        \cmidrule{2-6}
        & MCQ-DB & 882  & Accuracy & New & Unseen \\
        & MCQ-DA & 332  & Accuracy & New & Unseen \\
        & MCQ-ML & 556  & Accuracy & New & Unseen \\
        \midrule
        \multicolumn{3}{l}{\hspace{-.5em} \textit{\datadiscovery}} \\
        \cmidrule{2-6}
        & \birdselect     & 1,534   & Accuracy & Re-Purpose   & Seen \\
        & \begin{tabular}[l]{@{}l@{}} \texttt{Open-Wiki-} \\ \texttt{Table-TS} \end{tabular} & 5,134   & Accuracy & Re-Purpose  & Seen \\
        & \wikipage       & 104     & Accuracy & New  & Unseen \\
        \midrule
        \multicolumn{3}{l}{\hspace{-.5em} \textit{\texttosql}} \\
        \cmidrule{2-6}
        & Spider-dev & 1,034 & Exec. Acc. & Public & Seen \\
        & BIRD-dev   & 1,534 & Exec. Acc. & Public & Seen \\
        \bottomrule[1.2pt]
    \end{tabular}
  }
\end{table}

\para{Evaluation Datasets} Table~\ref{tbl:all_eval_dataset} provides an overview of evaluation benchmarks categorized into three main tasks: \mmlu, \datadiscovery, and \texttosql. It lists datasets used for evaluation, the number of examples in each, the metric applied, their data sources, and whether there are training sets included in the training corpus (i.e., in or out of distribution). We provide a detailed explanation of the data generation procedure in the next sections.

\para{Inference Sampling Hyperparameters}  We experiment with different sampling hyperparameters: temperatures of 0.0, 0.7, and 1.0, and top\_p values of 0.99 and 0.95. Lower temperatures yield more predictable responses, while higher temperatures encourage creativity. The top\_p value used in the nucleus sampling defines the range of tokens considered during generation, with higher values expanding the range. After tuning, we fix the hyperparameters for all subsequent experiments. For \mmlu, \datadiscovery and \texttosql, we set the temperature to be 0.0, 0.7, 1.0 and top\_p as 0.99, 0.95, 1.0, respectively.

\para{Inference and Model Serving} We use the vLLM model-serving framework~\citep{vllm} for inference. First, we convert the saved model checkpoints from Nemo format to Huggingface format using a conversion toolkit. We then deploy the Huggingface-formatted model using vLLM.

\begin{table*}[t!] 
 \centering
 \caption{The accuracy scores (\%) on \mmlu, \datadiscovery, and \texttosql tasks. "Unseen" datasets refer to those training sets that are not observed during \modelname post-training. "FE" (Format Error) indicates that the output answer format does not follow the prompt instructions. Results labeled “(FE)” include additional answer templates to improve extraction recall. "LCE" (Long Context Error) denotes the model's failure to comprehend lengthy input contexts. "CLE"(Context-Length-Exceeded) indicated that the input exceeds the model's maximum context length.} 
 \label{tbl:exp-main}
 \resizebox{\linewidth}{!}{
    \begin{tabular}{rr|ccc|c|| ccc|c || cc | c || c }
 \toprule[1.2pt]
 \multirow{5}{*}{Model} &  \multirow{5}{*}{Size}   & \multicolumn{4}{c||}{\mmlu} & \multicolumn{4}{c||}{\datadiscovery} & \multicolumn{3}{c||}{\texttosql} & \multirow{5}{*}{\begin{tabular}[c]{@{}c@{}} Over- \\ all \end{tabular}} \\  \cmidrule{3-13}
 & & \texttt{MCQ-DA} & \texttt{MCQ-DB} & \texttt{MCQ-ML} & \multirow{2}{*}{Avg.} & \begin{tabular}[c]{@{}c@{}} \texttt{BIRD} \\ \texttt{-TS} \end{tabular} & \begin{tabular}[c]{@{}c@{}} \texttt{Open-Wiki} \\ \texttt{Table-TS} \end{tabular}  & \begin{tabular}[c]{@{}c@{}} \texttt{WikiPage} \\ \texttt{-TS} \end{tabular}  & \multirow{2}{*}{Avg.} &  \begin{tabular}[c]{@{}c@{}} \texttt{Spider} \\ \texttt{-dev} \end{tabular}  & \begin{tabular}[c]{@{}c@{}} \texttt{BIRD} \\ \texttt{-dev} \end{tabular} & \multirow{2}{*}{Avg.} & \\ \cmidrule{3-5} \cmidrule{7-9} \cmidrule{11-12}
 & & Unseen & Unseen & Unseen &  & Seen & Seen & Unseen & & Seen & Seen & &  \\
 \midrule \midrule
Mistral-7B    & 7B  & 69.0 & 67.9 & 66.5 & 67.8   &  5.7 \tiny{(FE)} & 41.1 \tiny{(FE)}  & 28.8 \tiny{(FE)}  & 25.2   & 54.3 & 19.1 & 36.7  & 43.2 \\
Codestral     & 22B & 69.0 & 67.8 & 69.6 & 68.8   & 56.9 & 57.6 & 40.4 & 51.6  & 69.9 & 27.8 & 48.9  &  56.4\\
Mistral-Small & 22B & 71.1 & 74.3 & 70.9 & 72.1   & 48.6 & 66.3 & 25.0 & 46.6  & 18.6 & 5.8  & 12.2  & 43.6 \\
Mixtral-8x7B  & 47B & 75.0 & 72.9 & 70.7 & 72.9   & 35.8 & 2.3 \tiny{(FE)}  & 0.0 \tiny{(LCE)}  & 12.7  & 67.9 \tiny{(FE)} & 27.0 \tiny{(FE)} & 47.5 & 44.3 \\
 \midrule
GPT-3.5 Turbo & -   & 75.6 & 73.6 & 72.1 & 73.8   & 61.5 & 60.5 & CLE  & -     & 62.4 \tiny{(FE)} & 29.3 \tiny{(FE)} & 45.9  & - \\
GPT-4o-mini   & -   & 78.6 & 79.7 & 75.0 & 77.8   & 66.1 & 65.5 & 47.1 & 59.6  & 70.4 \tiny{(FE)} & 30.9 \tiny{(FE)} & 50.7  & 62.7 \\
GPT-4o        & -   & \textbf{80.1} & \textbf{82.9} & \textbf{80.0} & 81.0   & 71.7 & 71.1 & \textbf{58.7} & 67.2  & 71.7 \tiny{(FE)} & 34.4 \tiny{(FE)} & 53.1  &  67.1 \\
 \midrule
Mistral-NeMo & 12B & 68.7 & 71.5 & 72.7 & 71.0    & 41.2 & 29.3 & 28.8 \tiny{(FE)}  & 33.1  & 65.2 & 27.0 & 46.1 & 50.1\\
\modelname   & 12B & \textbf{77.1} & \textbf{77.6} & \textbf{74.3} & 76.3   & \textbf{\underline{78.2}} & \textbf{\underline{91.9}} & \textbf{55.8} & 75.3 &  \textbf{\underline{78.1}} & \textbf{\underline{37.1}} & 57.6 &  \textbf{\underline{69.7}} \\
 \midrule
 \multicolumn{2}{l}{ \textit{\footnotesize Improvement (\%) over base}} & 12.2\%$\uparrow$ & 8.5\%$\uparrow$ & 2.2\%$\uparrow$ & 7.6\%$\uparrow$ & 89.8\%$\uparrow$ & 213.7\%$\uparrow$ & 93.7\%$\uparrow$ & 127.5\%$\uparrow$ & 19.8\%$\uparrow$ &  37.4\%$\uparrow$ & 24.9\%$\uparrow$ & 39.3\%$\uparrow$\\
  \multicolumn{2}{l}{ \textit{\footnotesize Improvement (\%) over best}} & 3.7\%$\downarrow$ & 6.4\%$\downarrow$ & 7.1\%$\downarrow$ & 5.8\%$\downarrow$ & 9.1\%$\uparrow$ & 29.3\%$\uparrow$ & 4.9\%$\downarrow$ & 12.1\%$\uparrow$ & 8.9\%$\uparrow$ &  7.8\%$\uparrow$ & 8.6\%$\uparrow$ & 4.0\%$\uparrow$\\

 \bottomrule[1.2pt]
    \end{tabular}
  }
\end{table*}

\para{Results Overview}
Table~\ref{tbl:exp-main} presents the accuracy results across eight datasets, along with the average accuracy scores for each task. Notably, \modelname achieves the highest overall score of 0.697, surpassing the base model by 39.3\% and outperforming the best model (GPT-4o) by 4.0\%.

The evaluation consists of three tasks across eight datasets. Among these, the three \mmlu datasets are completely new. We do not explicitly include any multi-choice questions in the training corpus; however, the synthesized QA pairs in Chapter 1 may contain such question types. For the \datadiscovery task, we add the training sets of \birdselect and \openwikiselect into the training corpus as in Chapter 3 and evaluate on the test sets. Notably, we have not included any examples from \wikipage in the training corpus, make it a total new dataset for the post-trained model. Therefore, we mark \texttt{MCQ-DB}, \texttt{MCQ-DA}, \texttt{MCQ-ML}, and \wikipage as "unseen" datasets and others as "seen" datasets, where a subset of examples are included in the training corpus.
From Table~\ref{tbl:exp-main}, we find that  \modelname surpasses GPT-4o on all seen datasets and notably achieves a 213.7\% improvement over the base model on \openwikiselect. Analyzing the results on unseen datasets sheds light on the model's ability to generalize to new tasks or datasets. While \modelname does not always exceed GPT-4o's performance in this category, it nonetheless shows a significant boost over its base model. The largest relative improvement is observed on \wikipage, where it earned a 93.7\% enhancement.

\subsection{Analytics-specific Knowledge Testing}
\label{sec:analytics_mmlu}

First, we examine how well \modelname absorbs knowledge in the analytics field. Inspired by the commonly-used benchmark, Massive Multitask Language Understanding (MMLU)~\citep{hendrycks2020mmlu,wang2024mmlupro}, we curate a new dataset, ~\mmlu, to measure the model's capabilities in language understanding and reasoning in the analytics domain.

\para{New Datasets}
The \mmlu dataset consists of thousands of multiple-choice questions (MCQs) across three critical areas in analytics: database management (DB), data analysis (DA), and machine learning (ML). This results in three distinct datasets: \texttt{MCQ-DB}, \texttt{MCQ-DA}, and \texttt{MCQ-ML}. The questions feature complex queries that require models to exhibit deep expertise and advanced problem-solving abilities to achieve high accuracy. We source some of the questions from textbooks and generated additional questions and answers using \claudetf. All answers are manually reviewed and revised by three annotators to ensure quality. Table~\ref{tbl:all_eval_dataset} summarizes the data statistics. We use the \textit{accuracy} score to evaluate performance.

\para{Task Prompts}
The adopted prompt consists of a question, followed by four answer choices, and the required answer format. The task is to select the correct answer from the given choices. Here, we show an input-output example from \texttt{MCQ-DA}.
\begin{tcolorbox}[left=2pt, right=0pt, top=1pt, bottom=1pt]
\begin{verbatim}
# Input: 
You are an expert in data analytics. Answer the following MCQ. 
Question: Which of the following indicates no relationship in terms
of correlation?
Choices:
A: Cor(X, Y) = 1     B: Cor(X, Y) = 0
C: Cor(X, Y) = 2     D: All of the mentioned
Return your answer symbol (e.g., A, B, C, D) starting with "Answer:",
and give your explanation.

# Output: 
Answer: [B]
Evidence: Correlation is a statistical method that measures the
strength and direction of the relationship between pairs of variables.
\end{verbatim}
\end{tcolorbox}

\para{Main Results and Analysis} From Table~\ref{tbl:exp-main}, we note that \modelname, equipped with 12 billion parameters, consistently outperforms other open-source LLMs across all three categories of \mmlu, including \texttt{Mistral-7B}, \texttt{Codestral-22B}, \texttt{Mistral-Small-22B} and \texttt{Mixtral-8x7B}. Additionally, it surpasses \texttt{GPT-3.5-Turbo} but falls 1.5 and 4.7\% short of \texttt{GPT-4o-mini} and \texttt{GPT-4o}, respectively, in absolute accuracy. Compared to the base Mistral-NeMo model, \modelname achieves consistent performance improvements, with the largest relative gains of 12.2\% on MCQ-DA, 8.5\% on MCD-DB, and 2.2\% on MCQ-ML. This demonstrates the effectiveness of fine-tuning an LLM on diverse analytics-related QA tasks as curated in Section~\ref{sec:chapter1}.

We further conduct case studies to understand why \modelname can outperform the base model. Smaller models like Mistral-NeMo often need more specialized knowledge to answer specific knowledge-intensive questions. For example, Mistral-NeMo gives an incorrect answer and explanation for the example below. However, after being trained with the Chapter 1 data, \modelname gained more knowledge about E-R diagrams, enabling it to provide correct answers.

\begin{tcolorbox}[left=1pt, right=0pt, top=1pt, bottom=1pt]
\begin{verbatim}
Question: In the E-R diagram, generalization is represented by ____
A: Ellipse     B: Dashed ellipse
C: Rectangle   D: Triangle

# Predictions from Mistral-NeMo:
Answer: [A] 
Explanation: In an Entity-Relationship (E-R) diagram, generalization 
is represented by an ellipse (A). 

# Predictions from CoddLLM: 
Answer: [D]
Explanation: Generalization in E-R diagrams shows a superclass-
subclass relationship between entity types. It is depicted using an 
isosceles triangle pointing upwards, connecting the superclass 
(more general entity) to its subclasses (more specific entities).
\end{verbatim}
\end{tcolorbox}

\subsection{Table Selection}
Table selection aims to identify the most relevant subset of tables from a pool of candidate tables to answer a specific natural language question. Understanding how datasets complement or contradict each other helps users better determine which datasets provide the most relevant information. For example, related datasets may share common attributes or originate from similar domains, resulting in richer insights when combined. Conversely, recognizing discrepancies or redundancies among datasets can help void errors and misinterpretations in analysis. 

\para{Datasets} 
We have created three evaluation benchmarks to evaluate various scenarios for table selection (TS). The first two datasets are derived from BIRD~\citep{bird} and Open-WikiTable~\citep{open_wikitable}. The third dataset is our newly annotated benchmark, which includes text and table data, with carefully selected questions requiring rows from multiple tables. Specifically, \birdselect provides a controlled environment for assessing the model's performance on organized, relational data. \openwikiselect tests models' ability to discern and utilize subtle differences among similar tables. \wikipage tests the model's capacity for integrating information from various sources within a cohesive topic.
Table~\ref{tab: dataset_discovery} describes the statistics of the datasets. We mark \birdselect and \openwikiselect as seen datasets because the training corpus includes examples from the training sets of BIRD and Open-WikiTable as described in Section~\ref{sec:chapter3_table_selection}. \wikipage is an entirely new dataset primarily used for evaluation purposes.

\begin{table}[tb]
\caption{Summary of three evaluation datasets for \datadiscovery. "\#Cand. Table" stands for the average number of candidate tables per question. }
\resizebox{\linewidth}{!}
{
\begin{tabular}{lrccc}
\toprule[1.2pt]
\multicolumn{1}{l}{}   & \multicolumn{1}{c}{\begin{tabular}[c]{@{}c@{}} \textbf{\#Input}\\  \textbf{Token}\end{tabular}} & \multicolumn{1}{c}{\begin{tabular}[c]{@{}c@{}}\textbf{\#Cand.} \\ \textbf{Table}\end{tabular}} & {\begin{tabular}[c]{@{}c@{}} \textbf{Data}\\  \textbf{Characteristics}\end{tabular}} & \multicolumn{1}{c}{\begin{tabular}[c]{@{}c@{}}\textbf{\#Ground Truth}\\ \textbf{Table}\end{tabular}} \\ \midrule
\birdselect & 1.1K & 9.1 & Schema only & Multiple \\ \midrule
\begin{tabular}[c]{@{}l@{}}\texttt{Open-Wiki-}\\ \texttt{Table-TS}\end{tabular} & 2.3K & 10 & \begin{tabular}[c]{@{}r@{}}Metadata+\\ 3 sample rows\end{tabular} & Single \\ \midrule
\begin{tabular}[c]{@{}l@{}}\wikipage \end{tabular}  &  12.5K  & 21.6 & Text + Table  & Multiple \\ 
\bottomrule[1.2pt]
\end{tabular}
}
\label{tab: dataset_discovery}
\end{table}

\para{\birdselect: Candidate tables with well-designed schemas}
We employ a data processing method similar to that outlined in Section~\ref{sec:chapter3_table_selection} for the development data of BIRD, from which we select a total of 1,534 questions. For each question, we use all tables from the same database as candidate tables, relying solely on the well-designed schemas without using table content. The ground truth tables are derived from the corresponding gold SQL queries. It is important to note that some questions in this dataset require multiple tables to generate correct answers, highlighting the need for a deeper understanding of relational database design and the interdependencies between tables.

\para{\openwikiselect: Similar candidate tables with metadata and sample rows} For Open-WikiTable, we apply a similar method as described in Section~\ref{sec:chapter3_table_selection}, this time using the test set. We obtain 5,134 questions from the test set, ensuring ten candidate tables are retrieved via BM25, including the gold table. We also include metadata for the WikiTable as model input, such as page titles, section titles, and captions, when available. To enhance scalability when working with large tables, we limit three sampled rows per table to the LLMs. It is worth noting that each question in this dataset requires only one ground truth table. Therefore, we compare two retrieval-based methods: BM25~\citep{bm25} and an encoder-based method, \texttt{BGE-M3}~\citep{chen-etal-2024-m3}. Both methods learn the embeddings of the tables and questions, using cosine similarity to compute relevance scores. The table with the highest similarity score is then selected as the final table.

\para{Human-annotated \wikipage: Same Topic Text and Tables}
Existing Text-to-SQL and multi-table QA datasets~\citep{pal-etal-2023-multitabqa} consider table joins when generating programming languages or textual answers. However, these datasets assume that the key-foreign-key constraints (i.e., join relationships) are already provided. In data lakes, this assumption does not always hold. Moreover, it is more common to encounter both textual and tabular data in these contexts. 
To reflect this real-world scenario, we introduce a new human-annotated dataset that includes a Wikipedia page's data containing multiple tables and descriptions under the same topic.

\begin{figure}
    \centering
    \includegraphics[width=\linewidth]{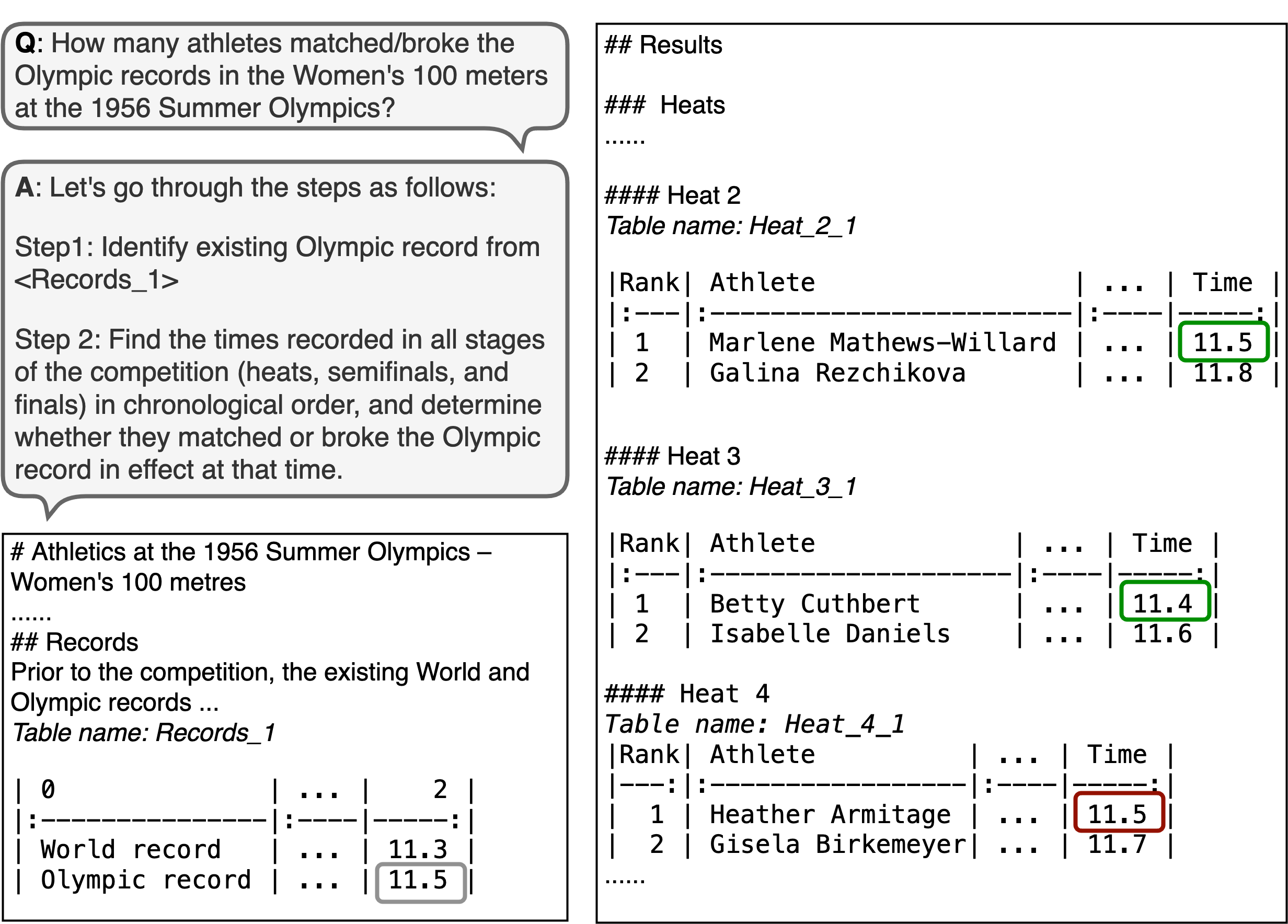}
    \caption{A question and wikipage data sample from \wikipage. In this example, we need to understand that \texttt{Heat\_2\_1}, \texttt{Heat\_3\_1}, and \texttt{Heat\_4\_1} occurred in chronological order. In \texttt{Heat\_2\_1}, Marlene ran 11.5 seconds, which matched the Olympic record. In \texttt{Heat\_3\_1}, Betty ran 11.4 seconds, breaking the Olympic record. By \texttt{Heat\_4\_1}, although Heather ran 11.5 seconds, the Olympic record had already been lowered to 11.4 seconds, so Heather did not match the new record. The answer is \textit{2 athletes}. For the \datadiscovery task, we regard all referenced tables as ground truth tables.}
    \label{fig:wikipage_heat}
\end{figure}

The data on the Wikipedia page contains various complex scenarios that require the model to possess both general knowledge and strong data interpretation capabilities. For example, in Figure~\ref{fig:wikipage_heat}, the question is: \textit{How many athletes matched or broke the Olympic records in the Women's 100 meters at the 1956 Summer Olympics?}. To answer this, we first identify the existing Olympic record from the \texttt{Record\_1} table, which is \textit{11.5 seconds}. The second step is to analyze the athletes' recorded times during the different stages of the competition (heats, semifinals, and finals) in chronological order to determine whether they matched or broke the record. In this example, the model needs to recognize that \texttt{Heat\_2\_1}, \texttt{Heat\_3\_1}, and \texttt{Heat\_4\_1} occurred sequentially. In \texttt{Heat\_2\_1}, Marlene ran 11.5 seconds, which matched the Olympic record. In \texttt{Heat\_3\_1}, Betty ran 11.4 seconds, breaking the Olympic record. By \texttt{Heat\_4\_1}, although Heather ran 11.5 seconds, the Olympic record had already been lowered to 11.4 seconds, so Heather did not match the new record. The correct answer is \textit{2 athletes}. If the model fails to grasp the definition of matching or breaking records, the chronological order of the heats, or that records can be broken at any point during the competition, it would struggle to provide the correct response.
The dataset creation and annotation steps are as follows:
\begin{enumerate}
    \item \parait{Step 1} We download approximately 100 Wikipedia pages related to the sports domain and convert them from HTML to plain text. The table data is converted into markdown format using \texttt{pandas.read\_html().to\_markdown()}. We name each table based on its section name, followed by a unique identifier. We flatten the structure of nested tables and omitted image data, links, or styling elements like bold text, borders, or colors. 
    \item \parait{Step 2} For each Wikipedia page, we use \claudetf to generate questions and identify the ground truth tables needed to answer them. We prompt the model to create questions requiring multi-table reasoning, ensuring that necessary information spans across multiple tables. Additionally, we instruct \claudetf to provide detailed step-by-step answers for each question. Although we do not explicitly use the answers, they serve as a reference for human annotators to more accurately evaluate the correctness of the reference tables in the next step. 
    \item \parait{Step 3} Three human annotators review and edit the examples to ensure the validity of the question and the accuracy of the ground truth tables (i.e., the labels). If we find any questions invalid, we return to step 2 to regenerate the question, reference tables, and answer triplets.
\end{enumerate}

Finally, we annotate 104 examples for the \wikipage dataset. On average, each question had 21.6 candidate tables, and the overall token count for a single Wikipedia page is approximately 12K, which is significantly longer than the other two datasets (Table~\ref{tab: dataset_discovery}).

\para{Evaluation Metrics and Baselines}
Unlike traditional dataset search settings that specify a predefined number of tables to retrieve, we allow the model to decide the number of necessary tables. We evaluate performance by using the \textit{accuracy} score, which measures the exact match between the predicted table set and the ground truth table set. The accuracy can be either 0 or 1. It equals 1 when the predicted set contains exactly the same tables as the ground truth set.
For \birdselect and \wikipage, the number of ground truth tables varies, making it challenging to apply traditional embedding-based methods to obtain the correct table sets. In contrast, for \openwikiselect, each question requires only one table, allowing us to use the top retrieved table as the prediction. Therefore, we include a lexical retrieval method, BM25~\citep{bm25}, and an encoder-based method, \texttt{BGE-M3}~\citep{chen-etal-2024-m3} as baselines.
 
\para{Task Prompts}\label{sec:table_selection_prompts}
We include the data information, the task instruction, and a question in the input prompt. For the data information, we use table schema for \birdselect, table metadata, three sample rows for \openwikiselect, and the entire Wikipedia page for \wikipage. These representations help tailor the data context for each dataset in the prompt. Here, we show an input-output example from \wikipage, which is based on the \href{https://en.wikipedia.org/wiki/100_metres_at_the_Olympics}{"100 metres at the Olympics"} Wikipedia page. To obtain the expected results for the question, we need first to identify the name of the male athlete who was the oldest champion in the \texttt{age\_1} table -- which is \emph{Linford Christie}. Then, we locate \emph{the year} when Linford Christie won the gold medal from the \texttt{Men\_1} table. 

\begin{tcolorbox}[left=1pt, right=0pt, top=1pt, bottom=1pt]
\begin{Verbatim}
# Input:
<Table Schema> / <Table Metadata>+<Sample Rows> / <Wikipedia Page>

Given the above data, identify one table or multiple tables that 
contain the necessary information to answer the following question.
\textbf{Question}: In which year did the oldest male champion win the 
100-meter Olympic games?
Provide the table name(s) within the <Tables> tag, with one table 
name per line.

# Output: 
<Tables>
Age_1
Men_1
</Tables>
\end{Verbatim}
\end{tcolorbox}

\begin{table}[t!]
\caption{The accuracy scores (\%) from \openwikiselect on two task-specific embedding-based baseline models.} 
\vspace{-3mm}
\label{tbl:open-wikitable-baseline}
\small
\begin{tabular}{r| rr |r}
 \toprule[1.2pt]
         & BM-25 & BGE-M3 & \modelname \\ \midrule
Accuracy & 71.1  & 77.8   & 91.9   \\
 \bottomrule[1.2pt]
\end{tabular}
\end{table}

\para{Main Results and Analysis}
\modelname demonstrates superior performance across all datasets. For the "seen" \birdselect and \texttt{Open- WikiTable-TS} datasets, where a subset of examples is used for training \modelname, performance improves by 89.8\% on \birdselect and 213.7\% on \openwikiselect compared to the base Mistral-NeMo model. Additionally, it outperforms the top model by 9.1\% on \birdselect and 29.3\% on \openwikiselect.

We observe a performance jump on the unseen \wikipage dataset, which has no examples included in the training phase. A key characteristic of this dataset is its long input context, averaging 12.5K tokens, with some examples exceeding \texttt{GPT-3.5-Turbo}'s maximum context length of 16,385 tokens. As a result, \texttt{GPT-3.5-Turbo} encounters Context Length Errors (CLE) during inference. The base \texttt{Mistral-Nemo} model fails on the \wikipage dataset because of its long input context and complex questions with table and text as input. Additionally, we observe that it fails to follow instructions and correctly output the required XML tag "<Tables>". To address this, we apply additional answer extraction techniques, treating the entire model output as a list of table names and interpreting each line as a separate table name. The refinement boosts accuracy from 0 to 0.288.  However, this result still falls far behind \modelname, which achieves 0.558 accuracy, representing a relative 93.7\% improvement over the base model. Results labeled "LCE" indicate long context errors, meaning the outputs are meaningless. Even after applying the above answer refinement technique, the results remain close to 0.

In comparison to embedding-based methods outlined in Table~\ref{tbl:open-wikitable-baseline}, where BM25 computes sparse representations, and BGE-M3 produces dense vectors, it is important to note that these baselines are applicable only to \openwikiselect since it is the only dataset with a single reference table. For this dataset, we use the top-1 predicted table as the final answer. However, for the other two datasets, where the ground truth labels contain multiple tables, embedding-based methods cannot handle such cases effectively. We observe that \modelname improves accuracy from 0.711 and 0.778 to 0.919, demonstrating superior data selection capabilities.

\subsection{\texttosql}
\label{sec:exp_text2sql}

\para{Datasets}
We evaluate model performance on the development sets of two public benchmarks, BIRD~\citep{bird} and Spider~\citep{yu-etal-2018-spider}, consisting of 1,534 and 1,034 examples, respectively.

\para{Evaluation Metrics} 
We use \textit{execution accuracy}~\cite{bird} as our primary evaluation metric. This measure evaluates whether the execution results of the predicted SQL query exact match those of the gold standard SQL query.  To compute execution accuracy, we adopt the evaluation script from the BIRD codebase.

\para{Task Prompts}  In this task, the model, acting as an SQL expert, receives the database schema and a question and outputs an appropriate SQL query to retrieve the information that answers the question.
To ensure a fair comparison of the model’s core capability in SQL generation, we adopt a standard zero-shot \texttosql prompt rather than optimizing for maximum accuracy with demonstrations, chain-of-thought reasoning, or data pre-processing techniques. Our zero-shot prompt consists of three main components: 
(1) \textit{Task instruction}: an initial prompt introducing the \texttosql task; 
(2) \textit{Data}: The table schema, including table names, column names, data types, as well as domain-specific knowledge when applicable (for the BIRD dataset only); 
and (3) \textit{Question}: the question for which a SQL statement needs to be generated. Note that we do not include any sample cell values in this prompt.

\begin{table*}[t!]
\caption{The accuracy scores (\%) on different model variants across all the datasets. We test two versions of Chapter 1 data: one with raw plain text and the other with instruction-response pairs derived from it. "Chapter1-Instructed+2+3" means that the model was first trained on the instructed version of Chapter 1 data and followed by training on Chapters 2 and 3. }
\vspace{-2mm}
\label{tbl:ablation}
\small
\begin{tabular}{r || ccc|| ccc|| cc}
\toprule[1.0pt]
 \multirow{4}{*}{\textbf{Model Variants}} & \multicolumn{3}{c||}{\mmlu} & \multicolumn{3}{c||}{\datadiscovery} & \multicolumn{2}{c}{\texttosql} \\  \cmidrule{2-9}
 & \texttt{MCQ-DA} & \texttt{MCQ-DB} & \texttt{MCQ-ML} & \begin{tabular}[c]{@{}c@{}} \texttt{BIRD-TS} \end{tabular} & \begin{tabular}[c]{@{}c@{}} \texttt{Open-Wiki} \\ \texttt{Table-TS} \end{tabular}  & \begin{tabular}[c]{@{}c@{}} \texttt{WikiPage-TS} \end{tabular}  &  \begin{tabular}[c]{@{}c@{}} \texttt{Spider-dev} \end{tabular}  & \begin{tabular}[c]{@{}c@{}} \texttt{BIRD-dev} \end{tabular} \\ \midrule \midrule
  Base Model & 68.7 & 71.5& 72.7 & 41.2 & 29.3 & 28.8 \tiny{(FE)} & 65.2 & 27.0 \\ \midrule
  Chapter1{\footnotesize -PlainText} & 65.2   & 68.6   & 63.5   & 4.6 \tiny{(FE)} & 0 \tiny{(LCE)} & 0 \tiny{(LCE)} & 6.5 \tiny{(FE)} & 26.7 \tiny{(FE)} \\
  Chapter1{\footnotesize -Instructed} & 68.1  & 73.2  & 71.4  & 50.1 & 80.5 \tiny{(FE)}  & 23.1 \tiny{(FE)} & 68.3 & 29.1  \\
  Chapter1{\footnotesize -Instructed}+3  & 70.5  & 72.0  & 72.1  & 78.0 & 89.9 & 46.2 & 77.2 & 35.6  \\ 
  Chapter2+3  & 73.5 & 73.6   & 73.7   & 79.3 & 92.8 & 53.8 & 74.2 & 33.8 \\
  Chapter1{\footnotesize -PlainText}+2+3 & 75.6  & 72.4  & 69.2  & 59.8 & 86.2 & 27.9 & 74.4  &  34.3  \\ \midrule
  CoddLLM -- Chapter1{\footnotesize -Instructed}+2+3  & \textbf{77.1}  & \textbf{77.6}  & \textbf{74.3}  & \textbf{78.2} & \textbf{91.9} & \textbf{55.8} & \textbf{78.1} & \textbf{37.1} \\
\bottomrule[1.0pt]
\end{tabular}
\end{table*}

\noindent An example prompt is provided below:
\begin{tcolorbox}[left=1pt, right=0pt, top=1pt, bottom=1pt]
\begin{Verbatim}
CREATE TABLE `Country` (
	`id`    INTEGER PRIMARY KEY AUTOINCREMENT,
	`name`  TEXT UNIQUE
) 
CREATE TABLE "Team" (
...

(Optional) -- External Knowledge: Perform better in crossing 
actions refers to MAX(crossing)

-- Using valid SQLite (and understanding External Knowledge), 
answer the following questions for the tables provided above.
# Question
Who are the top 5 players who perform better in crossing actions?
Generate the SQL within the <SQL> tag.
\end{Verbatim}
\end{tcolorbox}

We opt for this straightforward prompt to enable a fair comparison of the \texttosql capabilities without introducing complexities, such as multi-step reasoning. As noted by ~\citet{wretblad2024understandingeffectsnoisetexttosql} and ~\citet{kapoor2024aiagentsmatter}, such complexities can introduce unnecessary overhead, increase inference costs, and risk overfitting to specific dataset patterns. Nevertheless, as demonstrated in other tasks, \modelname effectively follows instructions and can be paired with complementary prompt engineering techniques when needed. 

\para{Main Results and Analysis} 
\modelname demonstrates strong performance in \texttosql, achieving an 8.9\% improvement over the zero-shot GPT-4o setting on Spider and a 7.8\% improvement on BIRD using the same prompt. Notably, GPT-4o, along with GPT-3.5-Turbo, GPT-4o-mini, and Mixtral-8x7B, struggled to adhere to the formatting instruction to "generate the SQL within the <SQL> tag". To address this, we apply an additional \verb|```sql```| code block template to extract more answers when the initial XML tag format fails. This augmented answer extraction resulted in execution accuracies of 0.717 on Spider and 0.344 on BIRD for GPT-4o.
Additionally, \modelname surpassed the base Mistral-Nemo model with improvements of 19.8\% on Spider and 37.4\% on BIRD.

Note that the accuracy scores in Table~\ref{tbl:exp-main} are not directly comparable to those on the public leaderboard. Our goal is to fairly assess the model's core capability in generating SQL queries rather than optimizing for higher scores across different datasets. We leave the exploration of using \modelname to generate more accurate SQL queries for future work. This includes providing demonstrations for in-context-learning~\citep{gao2024xiyan,pourreza2024chase}, generating multiple SQL candidates and then refining the best one~\citep{gao2024xiyan}, encouraging chain-of-thought reasoning~\citep{pourreza2024chase}, and applying schema pruning pre-processing methods~\citep{talaei2024chess}, which are used by leaderboard submissions.

Notably, BIRD is harder than Spider, involving more complex schema, queries that require the retrieval and joining of multiple tables, as well as incorporation of external knowledge. Our error analysis reveals that \modelname outperforms the base model by demonstrating (1) improved detection and generation of table joins using foreign key relationships,  
and (2) the ability to combine hints from the given external knowledge pieces with table schema.
For example, the following question from the BIRD dataset requires correctly joining the "schools" table with the "frpm" (an abbreviated name) table using the appropriate foreign key relationship. The base model fails to perform this task and overlooks the hint from the given domain knowledge. In contrast, \modelname generates a correct SQL query:

\begin{tcolorbox}[left=1pt, right=0pt, top=1pt, bottom=1pt]
\begin{Verbatim}
# Question
Please list the zip code of all the \textbf{charter schools} in Fresno 
County Office of Education.
# Domain Knowledge
\textbf{Charter schools} refers to \textbf{`Charter School (Y/N)` = 1} 
in the table frpm

# Prediction by base model
SELECT Zip FROM schools WHERE \textbf{Charter = 1} 
AND County = 'Fresno' AND District = 'Fresno County [...]]';

# Prediction by CoddLLM
SELECT DISTINCT s.Zip  FROM \textbf{schools s}
\textbf{JOIN frpm p ON s.CDSCode = p.CDSCode} 
WHERE s.District = 'Fresno County [...]' 
AND \textbf{p.`Charter School (Y/N)` = 1;}
\end{Verbatim}
\end{tcolorbox}

\subsection{Ablation Studies}

In this section, we conduct ablation studies to assess the impact of different training data formats and chapter selections. The accuracy scores of different model variants are presented in Table~\ref{tbl:ablation}. Note that \modelname was first trained on the instruction-response pairs of Chapter 1 data, followed by training on Chapter 2 and 3 datasets.

\subsubsection{Effects of the training on plain text documents v.s. Synthetic instruction-response pairs} To examine the impact of different data formats, we applied domain-adaptive \emph{continual pre-training}~\citep{ke2023continual} on plain text data and supervised instruction tuning on the instructed data version. Both models were built on an instructed model (Mistral-Nemo-Instruct) and trained using the cross-entropy loss for the next token prediction. However, during the continual pretraining process, the loss is calculated across all tokens to learn general language features and structures. For supervised instruction tuning, the loss is calculated only on the output (response) tokens, focusing on specific task objectives.

By comparing the accuracy scores of \texttt{Chapter1{\footnotesize -PlainText}} and \texttt{Chapter1{\footnotesize -Instructed}}, both trained exclusively on Chapter 1 data but with different data formats, we note that continual pre-training on plain text often degrades performance when applied to an instructed base model. On three \mmlu datasets, \texttt{Chapter1{\footnotesize -Instructed}} achieves scores of 0.681, 0.732, and 0.714, whereas \texttt{Chapter1{\footnotesize -PlainText}} scores lower at 0.652, 0.686, and 0.635. Additionally, on certain \datadiscovery and \texttosql datasets, the \texttt{Chapter1{\footnotesize -PlainText}} model completely fails, highlighting that training on domain text without clear task guidance can weaken the model's instruction-following capability. In contrast, \texttt{Chapter1{\footnotesize -Instructed}} matches or surpasses the base model in most datasets, suggesting that instruction-aware data more effectively aligns the model with the domain-specific queries and answers. 
Moving from \texttt{Chapter1{\footnotesize -PlainText}+2+3} to \modelname (\texttt{Chapter1{\footnotesize - Instructed}+2+3}) jumps \wikipage  by over 27 points (from 0.279 to 0.558) and boosts \mmlu and \texttosql tasks by several points.

\subsubsection{Effects of the Chapter 1 Analytics-specific Knowledge Corpus} 
Comparing the accuracy scores of the base model and  \texttt{Chapter1{\footnotesize -Ins- tructed}}, we observe that introducing Chapter 1 data leads to notable gains on domain-specific tasks while maintaining comparable performance on \mmlu. It improves accuracy on \datadiscovery and \texttosql, even without task-specific examples, demonstrating the effectiveness of instruction-aware domain adaptation. 
Comparing \texttt{Chapter2+3} and \modelname, we conclude that incorporating large-scale instructed instruction-response pairs from diverse tasks (Chapter 1) enhances overall model performance across a broad range of tasks, as evidenced by improvements in \mmlu. Furthermore, the instructed examples do not degrade the model’s performance on \datadiscovery and \texttosql, demonstrating its ability to generalize effectively without compromising domain-specific capabilities.

\subsubsection{Effects of the Chapter 2 Table and Text Alignment Data}
From Table~\ref{tbl:ablation}, by comparing \texttt{Chapter1{\footnotesize -Instructed}+3} and \modelname that includes all the three chapters data, chapter 2's table and text alignment data is crucial for tasks involving structured information. The highest surges in performance on table-centric discovery and \texttosql tasks occur when the model is supplemented with this alignment data. For instance, accuracy on \wikipage increases notably from 0.462 to 0.558 when Chapter 2 is added, showing how specialized alignment data about interpreting tables and text is instrumental for the model’s ability to handle more complex data discovery and retrieval tasks.

\section{Related Work}
\label{sec:related_work}

% This section positions \modelname in the context of LLMs for data analytics tasks and other domain-specific models.

\subsection{LLMs for Data Analytics}
LLMs are revolutionizing data management~\citep{wehrsteintowards} and analysis ~\citep{zhou2024llm}, and enabling new capabilities~\citep{fernandez2023large} across various tasks like data discovery~\citep{chorus}, metadata enrichment~\citep{nameguess,feuer2023archetype}, SQL query synthesis~\citep{zhang2024benchmarking}, and entity matching~\citep{zhang2024directions,peeters2023entity,zhang2024anymatch}.  
Several studies have employed "prompt engineering" techniques, carefully optimizing task instructions and few-shot examples to tackle specific tasks~\citep{llm_data_wrangling2022,peeters2023entity,chorus}. This approach is further enhanced by strategies like retrieval-augmented generation (RAG)~\citep{zhao2024chat2data}, which reduces hallucinations by incorporating domain-specific knowledge, and vector databases that enhance efficiency through semantic search capabilities~\citep{patel2024lotus}. 
LLM-based agents go beyond simple automation by LLMs to planning, writing code, executing it in an external sandbox, and interpreting results to solve complex data science~\citep{ds1000, dseval} and analysis challenges~\citep{InfiAgent-DABench,cao2024spider2}.
In addition to prompting LLMs without updating model weights, task-specific fine-tuning techniques have been widely adopted to boost performance on specific tasks~\citep{korini2024column,zhang2024anymatch}. 
Domain-specific fine-tuned models, like those for table understanding~\citep{table-gpt,su2024tablegpt2} and data wrangling~\citep{zhang2024directions, zhang2023jellyfish}, have been developed to enhance LLM performance across various tasks. 
Our proposed \modelname belongs to this category but stands out by leveraging a broader range of training tasks and examples. It focuses on understanding complex data relationships and specializes in natural language query-based tasks that require integrating data from multiple sources.

\subsection{Domain-specific Foundation Models}
Unlike training from scratch~\citep{wu2023bloomberggpt}, post-training is a commonly-used approach to build a domain-specific model across diverse fields such as mathematics~\citep{yue2024mammoth}, science~\citep{taylor2023galactica}, code~\citep{gunasekar2023textbooks}, finance~\citep{wu2023bloomberggpt}, and medicine~\citep{wu2024pmc}. Two
main approaches are continual domain-adaptive pretraining~\citep{mendieta2023towards,xie2023efficient} and instruction tuning~\citep{zhang2023instruction}. 
Continual pretraining involves training LLMs on large-scale, domain-specific text corpora. While this approach enriches the model with domain-specific knowledge, it may also degrade its ability to follow instructions effectively~\citep{ke2023continual}.
Instruction tuning applies the loss to well-prepared answers, necessitating a large number of input-output instruction pairs. 
MAmmoTH2~\citep{yue2024mammoth2} extracts instruction-response pairs from large-scale web corpus to enhance the model's reasoning capabilities. SciInstruct~\citep{zhangsciinstruct} employs a self-reflective instruction annotation approach to generate step-by-step reasoning for unlabeled scientific questions. Magicoder~\citep{wei2024magicoder} utilizes open-source code snippets to generate a diverse set of instructional data for coding tasks.
\modelname also applies instruction tuning to our well-curated, large analytics-specific training corpus. Unlike existing methods for creating instructional data, we leverage reference documents to synthesize large-scale, high-quality instruction-response pairs.
\section{Conclusion and and Future Work}
This work takes an initial step toward developing an expert analytics  model. We have taken the approach of curating a "textbook" that contains data to facilitate supervised instruction-tuning for a set of tasks that have not been considered in the past (e.g., \texttotable, \tabletotext, \datadiscovery), specifically in the context of foundation models for tabular data or analytics.  We argue that curating datasets and establishing benchmarks are crucial steps for advancing the next generation of LLMs for data analytics. While this is a laborious undertaking, it is an essential one. 
We believe that future work can focus on the following:

\para{RAG Systems and Tool Usage} Integrating retrieval-augmented generation (RAG) ~\citep{lewis2020rag} with \modelname to inject more fine-grained context or additional knowledge to the model is a promising direction. In addition, training the model to use advanced analytics tools~\citep{dseval, InfiAgent-DABench, datainterpreter} is crucial to tackle more complex tasks.

\para{Improved Training \& Evaluation Benchmarks} The human-annotated \wikipage benchmark is still in its early stages. While it is already useful to evaluate the current version of \modelname, the benchmark does not incorporate image or icon data found in Wiki pages. Second, we can  annotate the benchmark with step-wise supervision~\citep{lightman2024lets} for answering the questions. This type of annotations is useful to improve the model's reasoning ability. We are addressing the issues and planning a new release of the benchmark.

\newpage

\bibliographystyle{ACM-Reference-Format}
\bibliography{references}

\end{document}